# Patient-specific forecasting method for the functionality of a balloon-expandable stent

*Efstathios Stratakos[a], Vassilis Kostopoulos[a], Spyridon Psarras[a]\**,

[a]*Laboratory of Applied Mechanics, Mechanical Engineering and Aeronautics Department, University of Patras, Greece*
*University Campus 26504, Rio Achaia*

ARTICLE INFO

**ABSTRACT**

**In cardiovascular stenting procedures to treat Peripheral Artery Disease, idealized geometries fail to provide accurate data regarding the stent's long-term mechanical response. Therefore, image-based reconstruction methods of realistic incidents aim to develop a further perspicacity of the stenting process and counsel the physicians on the equipment decision-making. However, the stent placement simulation is still a complex problem due to high plastic deformations, multiple contact areas, and irregular geometries.**
**In this light, this research aims to develop a method with appreciable assumptions, reducing the time-cost and complexity of balloon-expandable stent simulations. A cardiovascular stent was constructed using stereoscopic images, while its mechanical response was vice-versa validated through Finite Element testing. A Computed Tomography of a patient with Critical Limb Ischemia was used to construct the 3D model of the curved common femoral artery. The mechanical response of the stent and the atherosclerotic artery was calculated using the Finite Element Method in a quasi-static analysis. Finally, the fatigue life cycles of the stent were estimated considering sinusoidal cyclic loading caused by the alteration of the systolic/diastolic pressure.**
**This research concludes that the crimping process of the stent is evaluated to have a negligible influence on its mechanical response. The recoil and the foreshortening percentages of the stent are predicted to remain static when increasing the number of rings of the stent, while a slight increase in stress concentration is observed. Finally, the numerical results indicate the biomechanical influence of stents placement in a realistic vessel, and comparisons are made with the literature. While the approach utilized reveals the feasibility of implementation on a wide range of cases, the prognosis of the stent's fatigue failure characterizes the placement of the stent as unsafe under the assumptions made.**

## 1. Introduction

Cardiovascular Diseases (CVDs) form a group of disorders of the heart and the blood vessels. Globally, CVDs are the leading culprit of death and serious illness, being responsible for an estimated 17.8 million lives in 2017, according to the American Heart Association [1]. The World Health Organization interprets this figure to 31% of deaths totality worldwide[2]. The number of deaths due to CVD is estimated to increase by at least one-third by 2030 [3]. Atherosclerosis is one of the most severe forms of CVD and constitutes the leading cause of death in the western world [4]. Peripheral Artery Disease (PAD) holds the third spot in terms of prevalence in the group of CVDs [5].

The usage of metallic stents to treat PAD has been one of the most ground-breaking and quickly implemented medical inventions of our times in the field of minimally invasive interventions [6]. As a minimally invasive method, implantation of stents to treat atherosclerosis includes a number of consequential assets over Endarterectomy, Bypass surgery, and Fibrinolytic therapy [7]. However, in-stent restenosis constitutes one of the major fetters of the stent's effort, which is caused by the sequence of the early thrombotic, the intermediate recruitment, and late neointimal proliferation [8,9].

Over the past few years, literature has substantiated that drug-eluting stents have reduced the restenosis rates compared to bare metal stents [10-13]. These stents are coated with an antiproliferative drug that inhibits the excessive growth of the neointima.

In our era, computational models using the Finite Element Method (FEM) have been widely adopted to study the angioplasty simulation. The results produced aim to offer an insight into the catheterization process, draw conclusions about the behavior of the implants and make suggestions on the material selections from the physicians and the manufacturing companies. In publications, idealistic geometries of the stent and the arterial tissue segment are constructed to study the behavior of the stent upon placement in the target vessel [14-19]. However, this approach may provide general directions regarding the procedure material selection and long-term performance of the stent, neglecting any imperfection of the geometries. Therefore, literature presents a necessity for geometries that depict accurately realistic models [20].

Under these requirements, researchers opt for image-based reconstruction models to investigate the angioplasty procedure using FEM [21-28]. Taking advantage of the studies' findings, the interventional cardiologists may make comparisons between the anatomical characteristics of the patient, aiming to select the optimum procedural planning. Patient-specific simulation models include increased time-cost, which has highly emerged on the imperfection of the 3D geometries and the combination of the ductile nature of the balloon-expandable (BE) stents along with the numerous contact condition alterations.

In this context, this work aimed to provide guidelines on the stenting simulation process establishing the feasibility of accurate and time-efficient patient-specific treatment. A reproduction of a patient's left Common Femoral Artery (CFA) suffering from Critical Limb Ischemia (CLI) has been developed using image-based reconstruction, which was reliant on the color-coding of the CT. The resulting surface model was manipulated in order to eliminate any irregularities that arose from the segmentation and reconstruction of the artery, engineering a fair replica. The atherosclerotic plaque and the artery were considered as two different bodies and their bonding mechanism was approached with 2 case studies. The healthy artery region was cut from the assemblage maintaining the proximate part to the formed atherosclerosis. Both models were regarded as homogeneous hyperelastic materials, nearly incompressible. Literature indicates that the irregularities of the stent generated at the manufacturing process should also be taken into consideration [20]. Thus, stereoscopic images from a manufactured stent were conjoined to construct the 3D geometry of the stent. Consequently, the constructed stent's mechanical response was evaluated through bench tests and compared with the manufacturing company's data. The inclusion of the manufacturing crimping process was evaluated among different discretization densities of the stent. Finally, using a time-step method the stent is expanded in the target vessel using an idealistic, hyperelastic cylinder as a balloon catheter. For the stent implantation, a case study request from the manufacturing company led to the selection of an undersized stent with respect to the length of the lesion. The fatigue life cycles of the stent implanted in the atherosclerotic vessel are calculated considering a sinusoidal strain alteration arising from the mean Systolic Blood Pressure (SBP) and Diastolic Blood Pressure (DBP).

\* *Corresponding author.* spsarras@upatras.gr

## 2. Materials and Methods

### 2.1. Arterial model segmentation from CT

The patient was a 64-years old female with Critical Limb Ischemia (CLI) and Type D lesions. The creation of the arterial model's CAD geometry was carried out through a process of CT segmentation, surface model creation, surface model treatment, solid model creation, and solid model treatment.

The segmentation of the target narrowed vessel from the rest of the body was conducted with the aid of the commercially available package 3D Slicer. The tomography was imported as a DICOM (Digital Imaging and communications in medicine) file [39] and after using volume rendering with the "MR-Angio" as a preset, the patient was depicted as shown in (Figure 1.A). The target vessel was found at the junction of the common iliac artery with the common femoral artery, illustrated in (Figure 1.B). The "Lookup Table" was set to "FullRainbow", with Interpolate ON and the Window/Level set manually to 1000/400. The Threshold was kept in the maximum span. The arterial wall, the arterial lumen, and the atherosclerotic plaque were visualized as discrete elements.

color coding shown in (Figure 1.E), with an accuracy of 0.0001 mm, reassuring the conformity of the constructed model. The formed model of surfaces for the arterial tree of the patient is depicted in (Figure 1.C). The area of the curved artery with the presence of atherosclerotic plaque was isolated. In a realistic model, the atherosclerotic plaque is embedded on the intima of the arterial wall and is not in direct contact with the arterial lumen. Instead, there is a thin layer of arterial tissue that acts as a boundary between the two elements. In relevant studies, this layer is neglected, and the atherosclerotic plaque is in direct contact with the arterial lumen [15,17-19,21,28,29,32,34,35,38]. This approach was also selected in the present study.

The three elements formed a discontinuity at their interface, indicated with the black circles in (Figure 1.D). This irregularity cannot be found in a realistic atherosclerotic artery. This issue was confronted by generating two different models of arteries.

In the first model, the arterial wall and the atherosclerotic plaque were discrete elements (Figure 2.A), while in the second model the atherosclerotic plaque was considered and declared as a mutual part of the arterial wall, (Figure 2.C). The atherosclerotic plaque is much stiffer than the arterial wall and different material properties had to be assigned to each of the two elements. The atherosclerotic plaque was eradicated from model 1 and the resulting artery is named model A (Figure 2.B), while the second model, considering the atherosclerotic plaque and the arterial wall united, was named model B (Figure 2.D).

These two surface models were imported in Spaiceclaim and converted into solid geometries. The solid geometries were inadequately meshed creating rough surfaces. The solid bodies were turned into "facet bodies" and smoothened, with a 0.01 mm gap size. Creating two perpendicular planes at the side cross-sections of each model, a skin was created on the surface of the bodies using 50 samples. Further smoothened surfaces were created and converted into solids. Through this process, the two models were simplified and correctified into a reduced number of easier-manageable surfaces. During this process, close attention was paid to the "deviation check", which is a graphical illustration that indicates that the processed bodies have an insignificant deviation from the original imported ones.

Model A and model B were then imported in a new assembly file, maintaining their initial X, Y, Z coordinates and the same coordinate system. The two bodies were overlapped. Model A represented the arterial wall and was completely covered by model B, which had a protruding part, representing the atherosclerotic plaque and and the covering inner, thin layer of the arterial tissue. The models were cut from one another at the intersecting edges. Three bodies were generated, the two of them being the double arterial wall and the third one representing the atherosclerotic plaque-inner arterial wall layer assembly. By eradicating one of the two identical arterial wall geometries, the artery-atherosclerotic plaque assembly was ready to be utilized for the FEA.

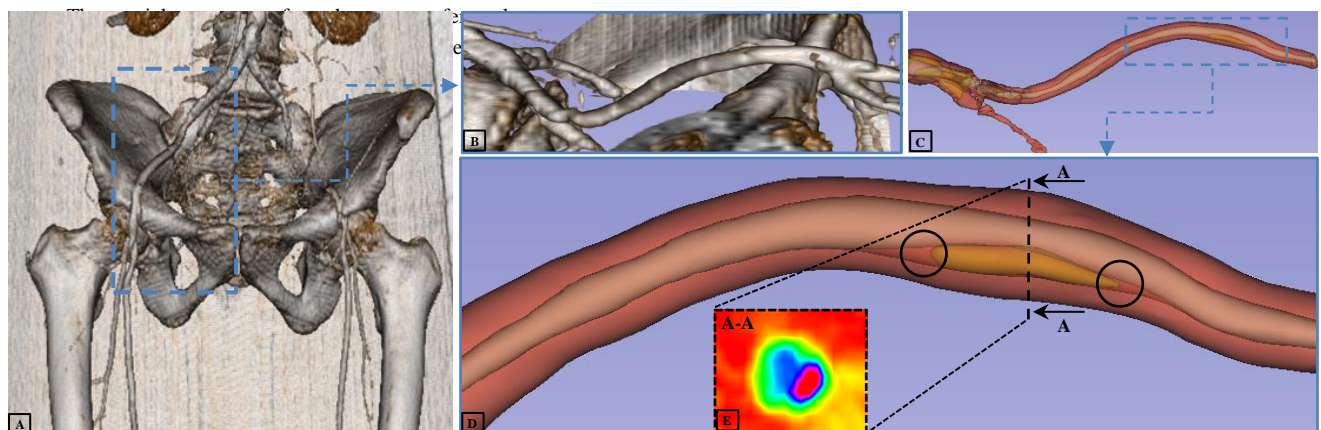

Figure 1. Segmentation of the arterial model using the program 3D Slicer: (A) Visualization of the CT of the 64-years-old patient, after being imported in 3D Slicer and having applied the MR-Angio filter (Coronal plane), (B) The curvature of the vessel in the area of interest at the left common femoral artery (Sagittal plane), (C) The model of surfaces for the arterial tree of the patient, after segmentation, (D) The vessel of the patient at the target area, where with the red colour illustrates the arterial wall, the azure represents the arterial lumen and the yellow shows the formed atherosclerotic plaque, (E) Artery cross-section indicating the arterial wall with the green colour, the arterial lumen with the azure, and the atherosclerotic plaque with the red colour.

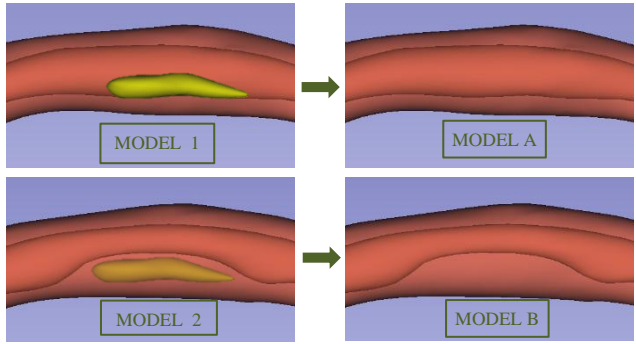

Figure 2. Contributing geometries to the final arterial model: (A) Model 1 includes both the atherosclerotic plaque with yellow color and the arterial wall with the red color, (B) Model A is Model 1 with the absence of the atherosclerotic plaque, (C) Model 2 considers the atherosclerotic plaque embedded in the arterial wall, (D) Model B is Model 2 with the absence of the atherosclerotic plaque

### 2.2. Stent and balloon angioplasty models

The stent that was utilized for the simulation was Zeus CC™, a BE stent manufactured by Rontis corporation. A number of bare-metal non-crimped stents were provided by the company. The stent's mesh was depicted by stereoscopic images, perpendicular to its cylindrical surface. The images were imported in the background interface of the commercially available package CATIA V5 and were assembled creating the 2D structure of the stents repetitive pattern. The pattern was projected on a cylinder with relevant dimensions to the original stent. The curved pattern was imported in ANSYS SpaceClaim and was transformed into a surface. The resulting surface was extruded into a solid pattern. Using a cylinder with 2 mm inner and 2.16 mm outer diameter, the solid pattern was subtracted from the cylinder. The cross-sectional corners were not rounded with fillets, as the ones found in the original geometry, to reduce the computational time-cost since curved edges would demand a denser mesh to keep up with the geometry.

The inner and outer diameter, the length, the cross-sectional width, and the thickness of the constructed stent were gauged. Next, the original stent was cut by a diamond cutting disk and polished at the edges. The resulting geometry was placed perpendicularly under the stereoscope. The stereoscopic images and the CAD constructed cross-section proved consistent. To further evaluate the validity of the stent as a replica of the real stent, its recoil and foreshortening percentages were calculated along with different numbers of rings and compared with the data provided by the manufacturing company.

The stent used for the expansion inside the arterial segment, considering its un-crimped configuration had a 2.16 mm of outer diameter, a strut thickness of 80 μm, and a length of roughly 200 mm.

Following the manufacturing process of balloon-catheter systems, the stent was crimped on an angioplasty balloon before being deployed. Therefore, a cylinder with 200 mm length, 22 mm external diameter, and 0.5 mm thickness was constructed, representing the crimping machine. The angioplasty balloon was simplified into a cylinder, with an external diameter of 17 mm, a 0.01 mm thickness, and a 220 mm length.

### 2.3. Material properties

Zeus CC™ consists of Cobalt-Chromium MP35N alloy and was considered isotropic. To define the material properties that were used as an input to the program, the company offered the stress-strain curve which accompanies each lot of stents. The curve was then translated to points using the program WebPlotDigitizer, as presented in Figure 3. These 54 produced points were introduced to describe adequately the elastoplastic behavior of the alloy, by using a Multilinear Isotropic Hardening model. The density was defined at 8430 kg/m$^3$ and Young's Modulus along with the Poisson's Ratio at 235 GPa and 0.3 respectively.

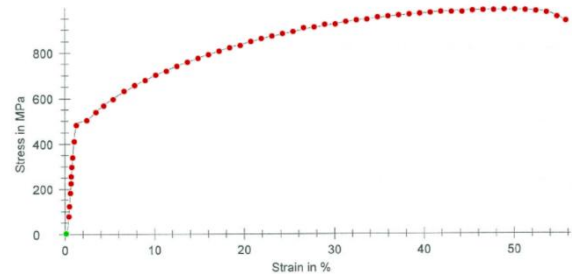

Figure 3. Stress-strain curve of Co-Cr MP35N translated into 54 points with coordinates of (Strain in %, Stress in MPa)

The mechanical properties of the arterial tissue and the atherosclerotic plaque of a specific patient can not be tested and assessed. The material properties may originate from relevant studies regarding the artery and the atherosclerotic plaque material characterization. Studies about simulations that investigate balloon-expandable stents in realistic artery models consider both elements as Mooney-Rivlin 5-parameter Hyperelastic models [29-34]. The material constants that describe the mechanical behavior of each element respectively are presented in Table 1.

| Property | Artery-value | plaque-value | Unit |
|---|---|---|---|
| Constant C10 | 0.018900 | -0.49596 | MPa |
| Constant C01 | 0.002750 | 0.51661 | MPa |
| Constant C20 | 0.059043 | 3.63780 | MPa |
| Constant C11 | 0.857200 | 1.19353 | MPa |
| Constant C30 | 0 | 4.73725 | MPa |

Table 1. Material properties of the artery and the atherosclerotic plaque

The Incompressibility Parameter $D_1$ is used to describe the compressibility of the material and is required as an input for the static structural simulation.

$$D_1 = \frac{2}{K}$$

where K is the bulk modulus [35].

In a study, 9 atherosclerotic arteries were mounted on a uniaxial tensile test machine, to investigate their material properties [35]. The incompressibility factor for the human atherosclerotic artery and the calcified plaque was calculated 0.045 and 0.174 respectively.

* *Corresponding author.* spsarras@upatras.gr

*2.4. Partitioning and discretization of models*

A high-quality and dense mesh is required to imprint the stent's geometric complexity and achieve accuracy in results. Artificial edges were interspersed to the stent's geometry to guide the elements desirably. Planes were introduced and manipulated to slice specific areas of the stent, slicing perpendicularly the length of the struts, at all 4 surfaces circumferentially of each cross-section. The junctions of the struts with the links were sliced with trapezoid-shaped planes to preserve the orderly flow of the mesh. The multizone mesh was utilized with mapped mesh controls for a higher-quality discretization. The multizone mesh was feasible to be implemented due to adequate and proper stent segmentation (Figure 4.C).

The larger stents were cut at the struts, producing a new body for each different ring. In the structural analysis, the intersections of mutual nodes were merged to maintain the coherency of the stent geometry. Using the quarter of a stent, a mesh convergence study was conducted in order to select the optimum grid for the stent. To constraint the stent, three frictionless supports were applied on the split surfaces of the stent, forcing the radial translation of the geometry, as illustrated in Figure 4.A.

To investigate the mesh convergence the maximum equivalent von Mises stress values were projected in an X-Y plot as a relation of different mesh densities (grid types), alternating the element size each time (Table 2). The stress values converged, and a deviation percentage bar chart was created to depict the deviation of each mesh type from the approximately correct stress value.

| Element No | 844 | 1386 | 2476 | 4204 | 15171 | 36599 | 46133 | 66312 |
|---|---|---|---|---|---|---|---|---|
| Grid type | 2x2 | 2x3 | 3x3 | 3x4 | 5x6 | 7x8 | 7x9 | 8x10 |
| Maximum Stress (MPa) | 826 | 850 | 854 | 872 | 887 | 893 | 897 | 899 |
| Deviation (%) | 8.1 | 5.4 | 4.97 | 5.94 | 1.34 | 0.62 | 0.2 | - |

Table 2. Mesh convergence study of the quarter stent

The restriction of relative movement between the atherosclerotic plaque and the arterial wall was implemented by setting mutual nodes at the interface. To ensure the existence of mutual nodes at the shared surface, the two models had to be discretized simultaneously. Shared topology is an adequate method to tie two surfaces together, preventing the breaking of their bond, a phenomenon that may occur due to large strain [40]. To guide the elements desirably, 6 planes were created by rotating each at a 30-degree angle from the previous. The intersection of these planes formed an axis parallel to the longitudinal axis of the arterial wall. On the grounds of the irregular geometry of the atherosclerotic plaque, the artery-plaque system was discretized with a Hex-Dominant method [36].

*2.5. Preliminary simulations*

Stents with different ring numbers were crimped and deployed. There were two groups of simulations, one that included and one that neglected the crimping process. Starting from a single-ring stent up to 16 rings, the maximum von Mises stress value was measured and compared with each group respectively. Regarding the stents with the full rings as components, a polar coordinate system was created at the center of the stent. Three vertices were constrained to move radially preventing their tangential and longitudinal movement (Figure 4.B). These vertices, having a 90-degree angle between them, restricted 2 degrees of freedom each, 6 DOFs in total. As a result, these constraints lead to an isostatic problem.

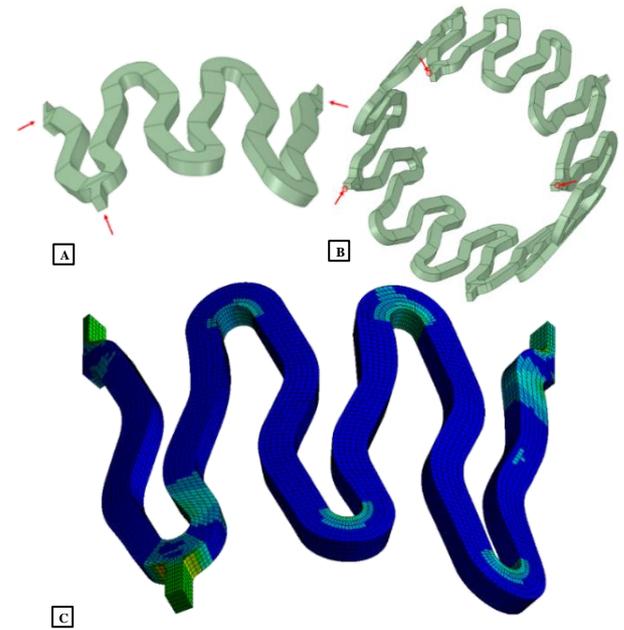

Figure 4. Boundary conditions on the stent: (A) Quarter of a ring indicating the 3 surfaces where frictionless support was applied, (B) Stent ring pointing the vertices that were constrained to move radially, (C) Discretization of the quarter-stent

*2.6. Deployment of the stent inside the artery*

The objective was the investigation of the behavior of the stent when deployed inside the atherosclerotic artery segment. to avoid mesh distortion between the stent and the artery-plaque assembly, two different mechanical models were created. These models were joined, introducing a third one, where boundary conditions between the models were set. The third model was imported in a static structural analysis. The normal pressure rates, following the norm of healthy females aging 61-65, for the Systolic Blood Pressure (SBP) account for 130.5 mmHg and the Diastolic Blood Pressure (DBP) 77.5 mmHg respectively [37]. The static structural simulation was conducted using 5 time-steps, which represented the stages shown in Table 3. To constraint the artery, 2 weak springs were placed on each side of the sliced arterial segment, in order to simulate the elastic force that the missing arterial segments create. The springs were perpendicular to each other and parallel to the sliced surfaces, with a body-ground scope.

Two cases of simulations were created. In the first case, shared topology was implemented at the whole mutual surface of the atherosclerotic plaque and the inner arterial wall, considering that the force needed to break the bond tends to infinity. In the second case, the shared topology constraint was implemented on a reduced surface area (Figure 5). It was assessed that the bond on the sides of the atherosclerotic plaque would break due to excessive force development.

After the accomplishment of the 5 steps, a fatigue tool was inserted to predict the fatigue life of the stent when alternating linearly between the step 4 and 5. The stress life of the stent was predicted through Goodman's mean stress theory [41].

| Step | Description | Loads | Contacts | Boundary conditions |
|------|-------------|-------|----------|---------------------|
| 1 | Application of the DBP at the arterial wall | • Pressure of 130.5 mmHg acting on the inner surface of the arterial wall | No contacts applied | • 2 weak springs on each side of the sliced artery<br>• 3 vertices constrained to move radially |
| 2 | Inflation of the angioplasty balloon | • Radial displacement of the angioplasty balloon from 1.97 mm to 5.4 mm | • Frictionless support angioplasty – balloon – stent<br>• Frictional support: stent – artery – plaque system | |
| 3 | Deflation of the angioplasty balloon | • Radial displacement of the angioplasty balloon from 5.4 mm to 3.4 mm | | |
| 4 | Alteration of DBP to SBP | • Reduction of pressure at the inner arterial wall surface from 130.5 mmHg to 77.5 mmHg | | |
| 5 | Alteration of SBP to DBP | • Increase of pressure at the inner arterial wall surface from 77.5 mmHg to 130.5 mmHg | | |

Table 3. Summary and description of the steps implemented on the expansion of the stent inside the artery simulation

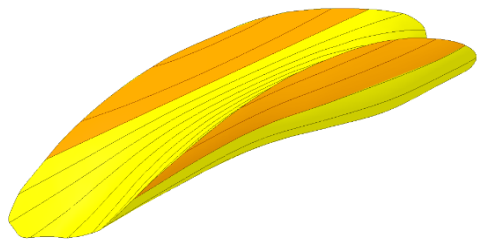

Figure 5. The atherosclerotic plaque where the yellow colour represents the shared topology area and the orange colour the unlinked area

## 3. Results and discussion

### 3.1. Preliminary simulations

In Figure 6, the influence of the crimping process on the stent residual stresses is examined. It was calculated that there is an average of 3.28% augmentation in stress rates when the crimping has occurred before the deflation of the balloon. It was also estimated that the crimping has an even more insignificant effect on the stent's structural behavior after the deflation of the stent, accounting for 2.08%. Therefore, the effect of the stent's crimping on the balloon catheter was considered negligible and the process was omitted from the final simulation.

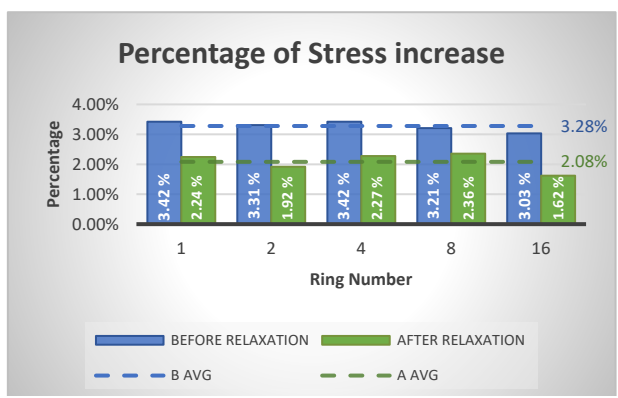

Figure 6. Percentage of stress concentration increase when including the crimping process in the as part of the simulation: a) Before relaxation (at the maximum expansion of the stent) and b) After relaxation (After the strain release on the stent)

The recoil and foreshortening proportions appeared relatively changeless when increasing the number of the stent's rings. The average recoil percentage was calculated to be 4.46% when the manufacturing company claims 4.6%. The respective foreshortening was assessed at 4.53%, where 4.5% is taken from the company's data. A slight increase in stress concentration was observed when the number of rings increases. This is justified by the torsion effects, occurring due to the stent's asymmetrical nature, which generates bending loads Figure 7.

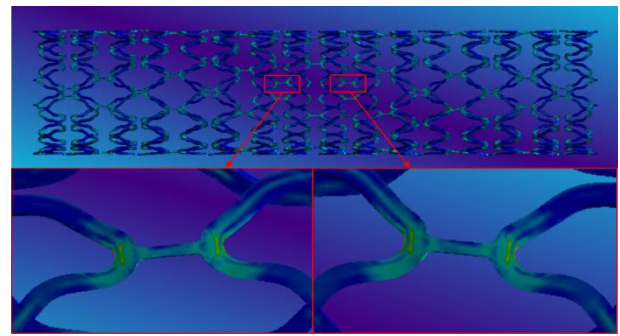

Figure 7. Torsion effect occurring between the different rings increasing the stress concentration rates

### 3.2. Complete surface shared topology

This simulation was performed to investigate the behavior of the stent when being deployed inside the artery, where the whole intersecting surface of the artery was considered mutual with the atherosclerotic plaque's bottom surface. After the expansion of the balloon, as the balloon retracts, the stent gets compressed by the arterial wall, as illustrated in Figure 8.

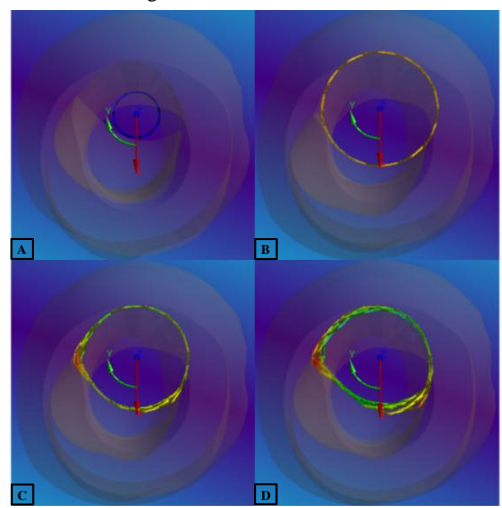

Figure 8. Cross-section of the directional deformation of the stent A) at 0 sec, B) at 1 sec, C) at 1.5 sec, D) at 2 sec – Polar Coordinate System X-Y axis, whole surface shared topology

* Corresponding author. spsarras@upatras.gr

The stent is crushed severely, mostly on the left side by the artery-plaque assembly, as well as slightly on the right side. However, this does not prove the structural inadequacy of the stent.

By observing Figure 9, there can be distinguished a high-stress concentration at the side edges of the atherosclerotic plaque, hitting a high of 17.4 MPa at 1 sec and dropping to 4.7 at 2 sec. These figures are significantly high taking into consideration that pulsatile pressure inside the vessel ranges from 0.01 MPa to 0.017 MPa. The justification behind the resulting stresses relates to the shared topology hypothesis. Since the side edges of the plaque are considered mutual, it is normal for high stress to develop at the interface between two materials with different stiffness values.

In reality, the bond between the artery and the atherosclerotic plaque would break if the stress gets high enough, exceeding the strength of the bond between the tissues. As a result, providing that there was data about the strength of the bond, the tearing at the interface could be modeled with Cohesive Zone Elements. However, no data was found available on the strength of this bond and this applies to the approaching method.

To conclude, this approach led to unrealistic results due to the assumption of the mutual nodes at the intersecting surfaces of the artery-plaque assemblage. In a real case scenario, the stresses surpassing the 1 or 2 MPa could lead to significant trauma of the arterial vessel. Thus, the strategy of implementing shared topology at the whole intersecting surface could be considered misleading.

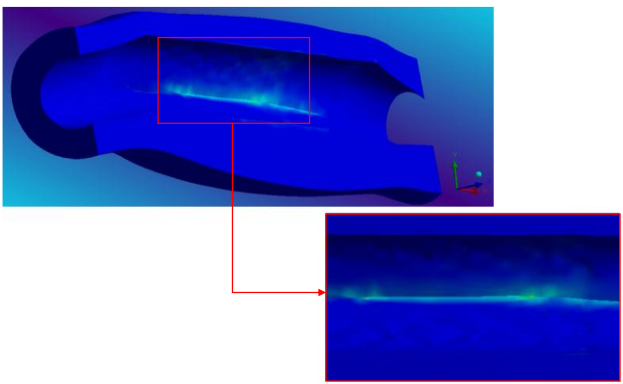

Figure 9. Equivalent Von Mises stress distribution of the artery-plaque assembly at 1 sec – sliced in the longitudinal axis

### 3.3. Reduced surface shared topology

This simulation represents the second approach of the stent expansion inside the atherosclerotic artery. In this case, the bond between the artery and the atherosclerotic plaque is considered unbreakable in a diminished area of contact. At 5 sec, the minimum inner diameter of the arterial wall was 4.92 mm, while the maximum was 5.26 mm. The goal was for the resulting artery to have a minimum 5 mm diameter lumen, which even though was not accomplished, was sufficiently approached. The stent was successfully expanded to an outer diameter of 5.4 mm and contracted indicating a recoil percentage ranging from 2.6% on the sides to 8.9% in the narrowest area.

In Figure 10, the directional deformation distribution of the stent is illustrated at the implementation of pressure in the artery, the expansion of the stent, and the deflation of the balloon. The stent is significantly less compressed on the left and right side at the end of the deflation, maintaining a greater lumen diameter of the artery. In addition, the stent is symmetrically deformed at the end of the deflation, which implies that the boundary conditions are properly set and there were no unwanted noticeable bending or torsional strains.

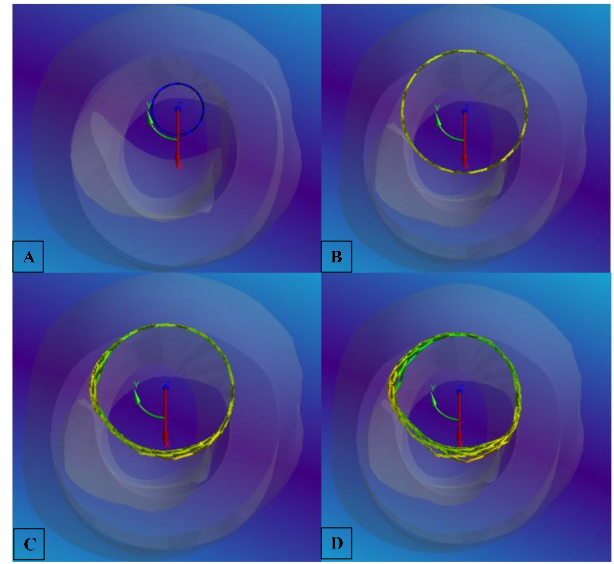

Figure 10. Cross-section of the directional deformation of the stent A) at 0 sec, B) at 1 sec, C) at 1.5 sec, D) at 2 sec – Polar Coordinate System X-Y axis - Reduced surface shared topology

In Figure 11, the Equivalent von Mises stress distribution at the stent throughout 3 different time-steps is presented. At 2 sec, the artery-plaque system is deformed according to the balloon-stent system geometry, while at 3 sec the stent has conformed to the vessel geometry.

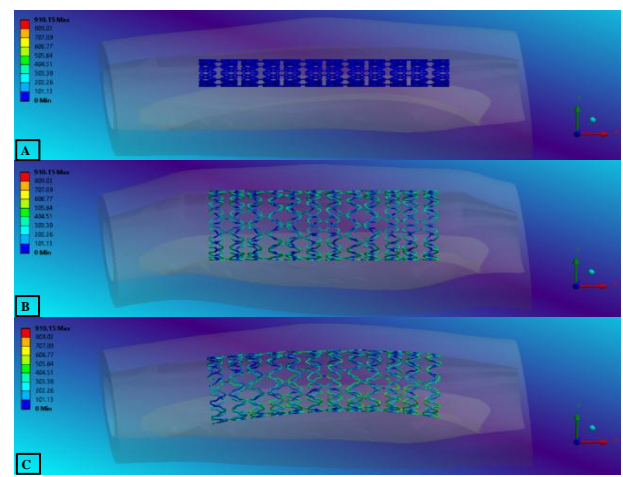

Figure 11. Equivalent Von mises stent distribution among the stent rings at A) 1 sec, B) 2 sec and C) 5 sec – Reduced surface shared topology

The maximum Equivalent von Mises stress on the stent is represented in detail in Figure 12. The stent indicates zero stresses applied on the first step, while there is a linear increase in stress value until the 1.2 sec, where the yield strength is exceeded, hitting a peak at 2 sec, when the maximum dilation of the balloon is accomplished. Next, as the balloon deflates, the stent releases its remaining elastic strains, until 2.1 sec when the stent would maintain its shape if the artery-plaque system was not present because the elastic strains of the vessel force the stent to further retract, augmenting its stress values. At 2.2 sec, the stent starts to resist its compression and the stress values remain relatively static. At 2.6 sec the phenomenon of stent recoil ends and until 3 sec, the stress values remain changeless. From 3 to 4 sec, as the pressure in the inner arterial wall decreases, the stress values indicate a slight growth, which is terminated with the pressure increase. In this last step, some of the remaining stresses that occurred in the precedent step are released but a greater percentage of them are translated into plastic strains.

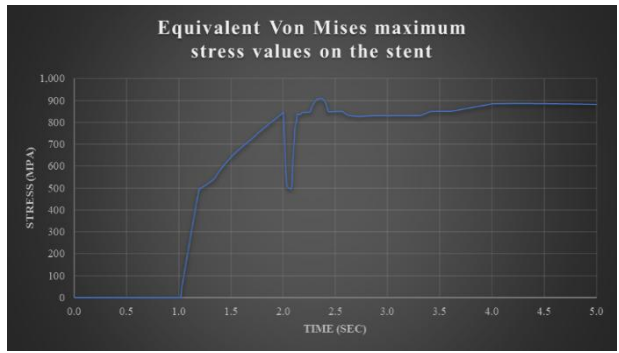

Figure 12. Maximum Equivalent von Mises stress values on the stent over time– Reduced surface shared topology

The critical stress areas are illustrated with red color in Figure 14, which takes place at the greater curvature of the pattern and specifically at the one that is forced to open the most. The fatigue life prediction of the stent is estimated according to the stress difference between the systolic and diastolic pressure. To calculate this result, a fatigue tool was inserted with default parameters. The 85,756,000 number of cycles is considered the endurance limit of the material. Theoretically, the component could be cycled at stress ranges below this level for an infinite number of cycles and it will never fail because of fatigue. The stent indicates failure at 45,400,000 cycles, while with red color the areas where the crack will start are indicated. This fracture may have a lot of justifications. The combination of arterial curvature, a great percentage of artery blockage due to plaque, and a reduced number of rings may be responsible for lower stent integrity. The construction of both the stent and the artery-plaque assembly may have geometrical imperfections that do not correspond with reality.

The stress distribution on the artery and the atherosclerotic plaque appears in Figure 13 and Figure 15 respectively. The stress values indicate a concentration at the edges of the artery-plaque bond and as aforementioned, this is explained by the difference in stiffness among the 2 materials. The mean stress at the contact between the artery and the stent is around 0.4 MPa, while the respective figure for the stent-plaque contact is 0.25 MPa. These figures are in accordance with relevant literature [38]. The arterial wall reveals greater strain numbers when comparing to those of the plaque due to

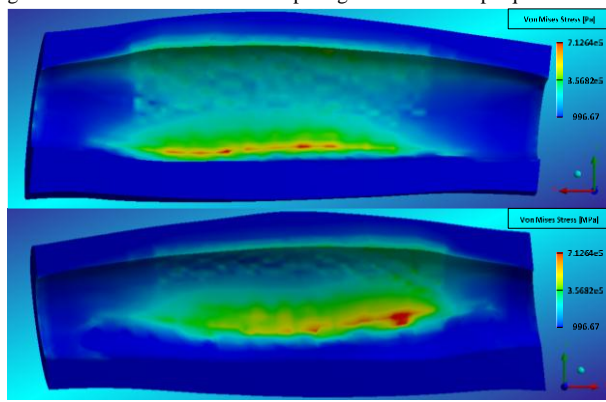

Figure 13. Equivalent Von Mises stress on the arterial wall– Reduced surface shared topology

*Corresponding author.* spsarras@upatras.gr

their compressibility difference since the plaque is much stiffer than the artery.

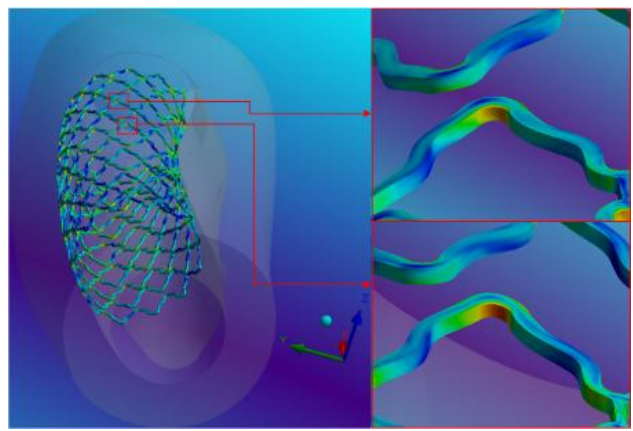

Figure 14. Fatigue life prediction of the stent indicating the areas where the failure is estimated to occur– Reduced surface

### 3.4. Study limitations and further developments

The experimental testing and result comparison is required to confirm the results of this study and of course necessary before the medical community takes into consideration this finite element analysis. This could be done by constructing an arterial segment with the same behavior in an electro-spinning machine and 3D printing the atherosclerotic plaque. To produce a point-to-point replication of the arterial wall's geometry, the lumen of the artery-plaque system should be 3D printed as well and specifically with a liquid dissolving material. The 3D printed plaque and lumen could be glued with dissolving glue, and the resulting geometry should be used as an axis to wrap around the fibers of the electro-spinner. After that, the lumen-plaque-artery system would be left in a tab with liquid they would dissolve the lumen as well as the lumen-plaque bond. Therefore, the resulting geometry would be similar to the one utilized in the finite element analysis.

To test the deployment of the stent, devices that can calculate micro-strains should be involved in order to estimate the stent remaining stresses. Then, the artery-plaque-stent system could be inserted in a pulsatile flow machine or an alternating pressure machine to gauge the life cycles of the stent under the alternating pressure.

Regarding the simulation, most materials used in this study were tested and validated. However, there are certain assumptions that were taken. The first one is the reduced number of stent rings that could have led to inaccurate results that do not represent reality. In future research, a stent that covers the whole surface of the plaque could be used with a denser mesh, to better capture the stress values on the stent. In addition, the deployment mechanism of the balloon should be the application of pressure for more accuracy, and specifically, the one that is suggested by the manufacturers. Besides, in the simulation, there could be an addition of pulsatile blood flow to investigate the influence that it may have on the stent's structural integrity.

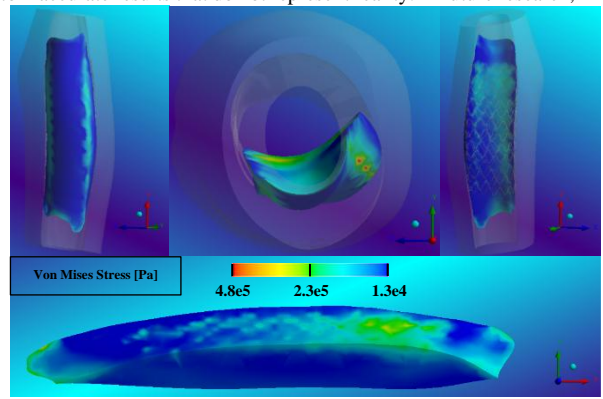

Figure 15. Equivalent Von Mises stress on the atherosclerotic plaque– Reduced surface shared topology

Finally, the bonding between the atherosclerotic plaque and the artery should be further studied since there was no data found available in the bibliography. Furthermore, when these data are available, a simulation with Cohesive zone elements should be conducted to study the influence of the bond on the deployment process. Further analysis in dynamic loading is required to draw a conclusion about how safe is it to place balloon-expandable stents on curved vessels since research indicates that stents that are not vulnerable to corrosion under static loading may fail and dynamic loading.

## 4. Conclusions

This study presents the feasibility of implementing patient-specific simulation in stenting procedures with limited computational and time costs. Detail guidelines were introduced regarding the artery segmentation from the CT and its 3D model construction, as well as the stent's replica formation. The models were validated and meshed sufficiently. Regarding the FEA, certain assumptions were made which led to stress magnitudes comparable to the preexisting literature. The necessity of the atherosclerotic plaque's adhesion on the arterial wall was exhibited, while two approaching methods were examined. The stent expanded the target vessel adequately when considering mutual nodes between the artery and the atherosclerotic plaque at a reduced surface. However, the fatigue life results indicate a great deviation from the required stent's functionality, implying deficient or erroneous assumptions. To conclude, this work encourages the patient-specific approach, while indicating the limitations caused by the complexity of the procedure.

ACKNOWLEDGMENTS

The authors thank MD. Panagiotis Kitrou, University of Patras, Greece for his scientific support in the medical aspects of the work as well as the patient selection for the case study.
The publication of the article in OA mode was financially supported by HEAL-Link

\* *Corresponding author.* spsarras@upatras.gr